\documentclass{article}
\usepackage{spconf,amsmath,graphicx,mathtools}
\usepackage{amssymb}
\usepackage[dvipsnames]{xcolor}
\usepackage{cite}
\usepackage{enumitem}
\usepackage{flushend}

\usepackage[font=small,skip=4pt]{caption}
\newcommand{\phZ}{\phantom{0}}
\newcommand\norm[1]{\left\lVert#1\right\rVert}

\title{Convolutional Block Design for Learned Fractional Downsampling}
\name{Li-Heng Chen$^{\star}$ \qquad Christos G. Bampis$^{\dagger}$ \qquad Zhi Li$^{\dagger}$ \qquad Chao Chen$^{\dagger}$ \qquad Alan C. Bovik$^{\star}$}
\address{$^{\star}$ Laboratory for Image and Video Engineering (LIVE), The University of Texas at Austin \\ $^{\dagger}$ Netflix, Inc.}

\begin{document}
\setlength{\abovedisplayskip}{5pt}
\setlength{\belowdisplayskip}{5pt}
\setlength{\abovedisplayshortskip}{2pt}
\setlength{\belowdisplayshortskip}{2pt}
%
\maketitle
\begin{abstract}
The layers of convolutional neural networks (CNNs) can be used to alter the resolution of their inputs, but the scaling factors are limited to integer values. However, in many image and video processing applications, the ability to resize by a fractional factor would be advantageous. One example is conversion between resolutions standardized for video compression, such as from 1080p to 720p. To solve this problem, we propose an alternative building block, formulated as a conventional convolutional layer followed by a differentiable resizer. More concretely, the convolutional layer preserves the resolution of the input, while the resizing operation is fully handled by the resizer. In this way, any CNN architecture can be adapted for non-integer resizing. As an application, we replace the resizing convolutional layer of a modern deep downsampling model by the proposed building block, and apply it to an adaptive bitrate video streaming scenario. Our experimental results show that an improvement in coding efficiency over the conventional Lanczos algorithm is attained, in terms of PSNR, SSIM, and VMAF on test videos.
\end{abstract}
\begin{keywords}
convolutional neural networks, downsampling, adaptive video streaming, perceptual video quality.
\end{keywords}
\section{Introduction}
\label{sec:intro}

Video signals have increasingly dominated the Internet in recent years, driven by the evolution of consumer electronics and the tremendous popularity of video sharing and streaming platforms. In a streaming video workflow, each component plays an important role in the end-to-end efficiency. For example, raw video sources directly determine the base video quality, while the selection of encoding parameters or rate control algorithms affect rate-distortion tradeoffs. An important recent advance is adaptive streaming framework, a technique that is widely used by video streaming services like Netflix, Youtube, or Facebook to improve the quality of experience of viewers \cite{CChen2018, KatsavounidisDO18, Wu2020}. 

\begin{figure}[!t]
  \centering
  \footnotesize
  \includegraphics[width=3.1in]{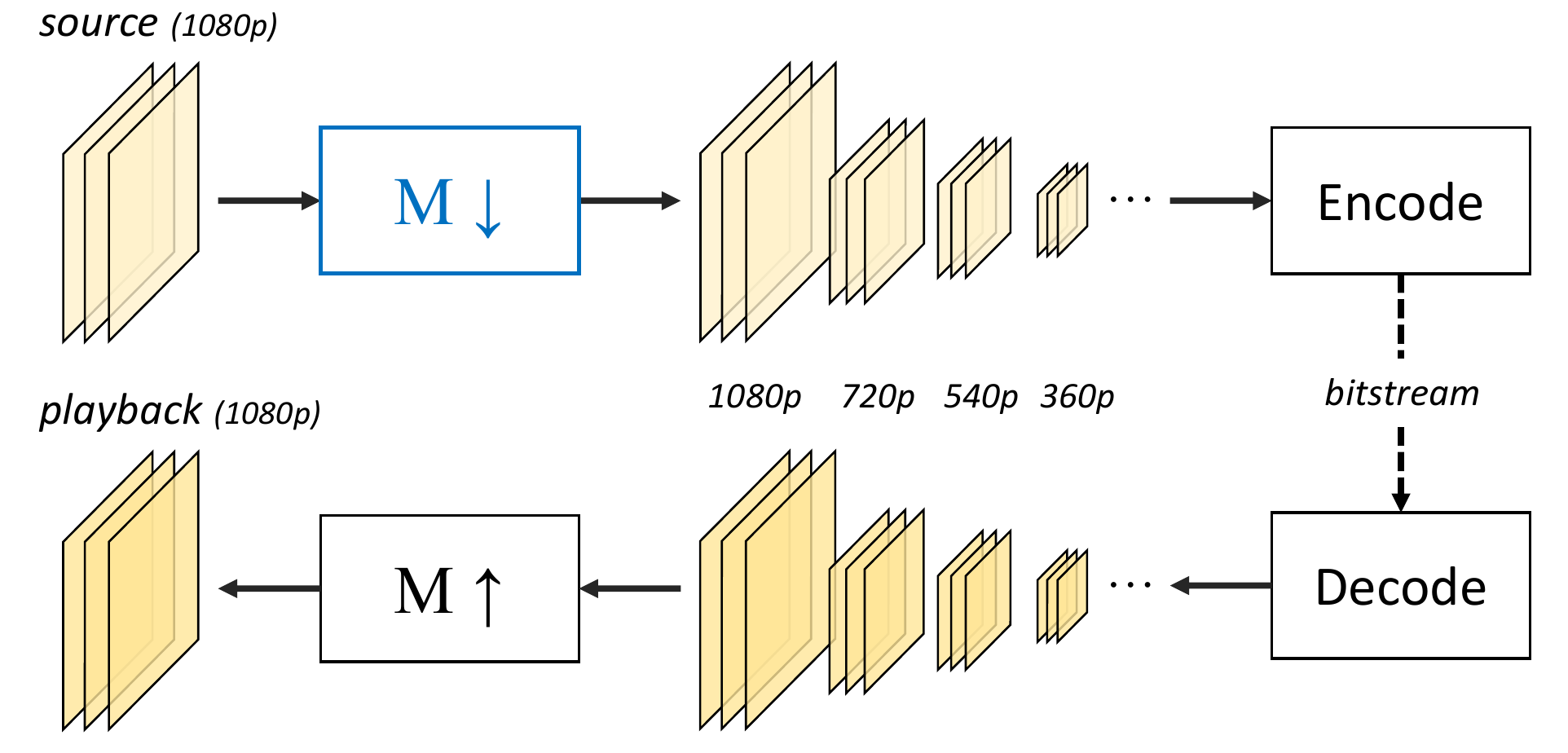}
  \caption{A general flow diagram of adaptive video streaming. The parameter $\mathit{M}\ge1$ denotes the current scaling factor, which may be varied. The downsampling block highlighted in blue is where our ideas are applied.}
  \label{fig_abr_flow}
\end{figure}

As shown in the workflow illustration in Fig. \ref{fig_abr_flow}, a source video segment (typically a scene) is downscaled into multiple resolutions using scaling factors $\mathit{M}\in\mathbb{Q}_{+}$. The videos at each resolution are then encoded using different quantization parameters (QPs), yielding a variety of rate-quality tradeoffs. From amongst the generated resolution-bitrate representations, a ``best'' video chunk is determined (typically by perceptual optimization), then streamed \cite{Toni2015, Li2016, Sani2017}. On the client side, the bitstream is decoded and scaled back to the device resolution prior to display. It is important to understand that the scaling factor $\mathit{M}\ge1$ may be any reasonable rational number. For instance, streaming a 1080p source at 720p resolution ($\mathit{M}=1.5$) is one of the most common choices for streaming services. The upscaling algorithms implemented on different display devices generally vary, and are not known by the stream providers. Hence, the encoding pipeline cannot be optimized for a specific upsampling algorithm.

\begin{figure*}[htb]
  \footnotesize
 \begin{minipage}[b]{0.33\linewidth}
   \centering
   \centerline{\includegraphics[width=5.9cm]{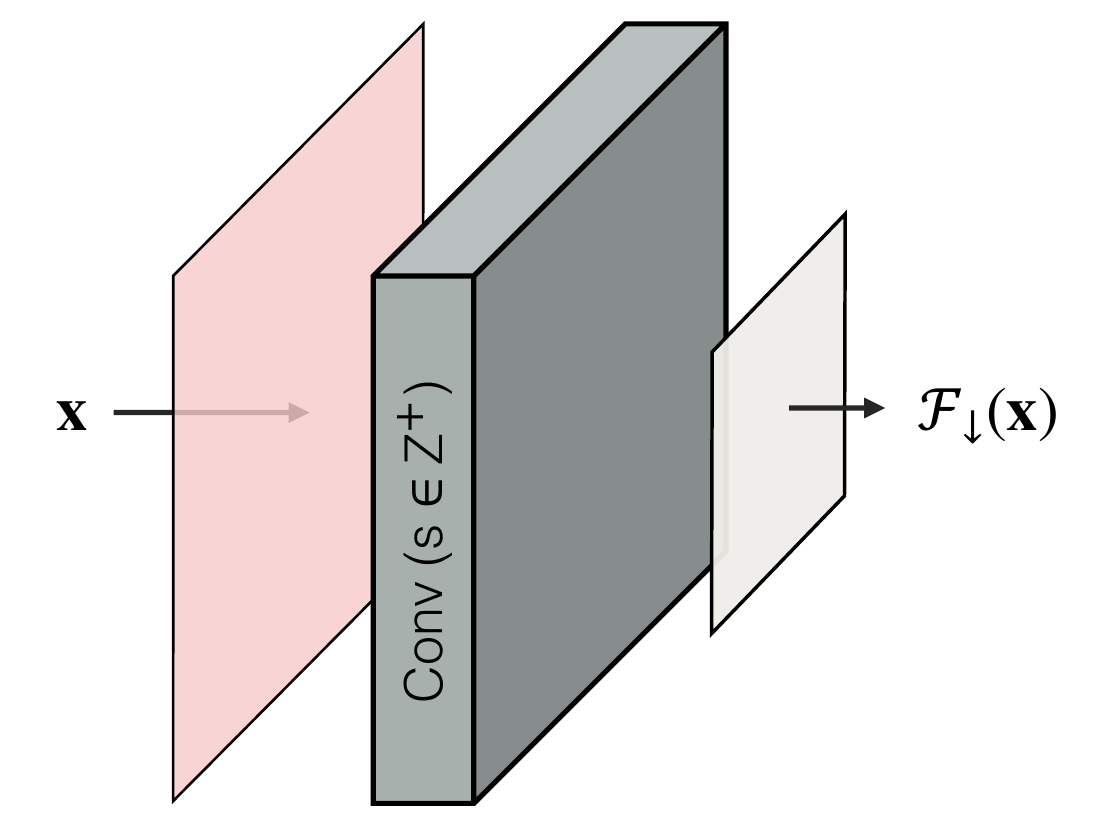}}
   \centerline{(a) CNN (integer $s>1$)}\medskip
 \end{minipage}
 \hfill
 \begin{minipage}[b]{0.33\linewidth}
   \centering
   \centerline{\includegraphics[width=5.9cm]{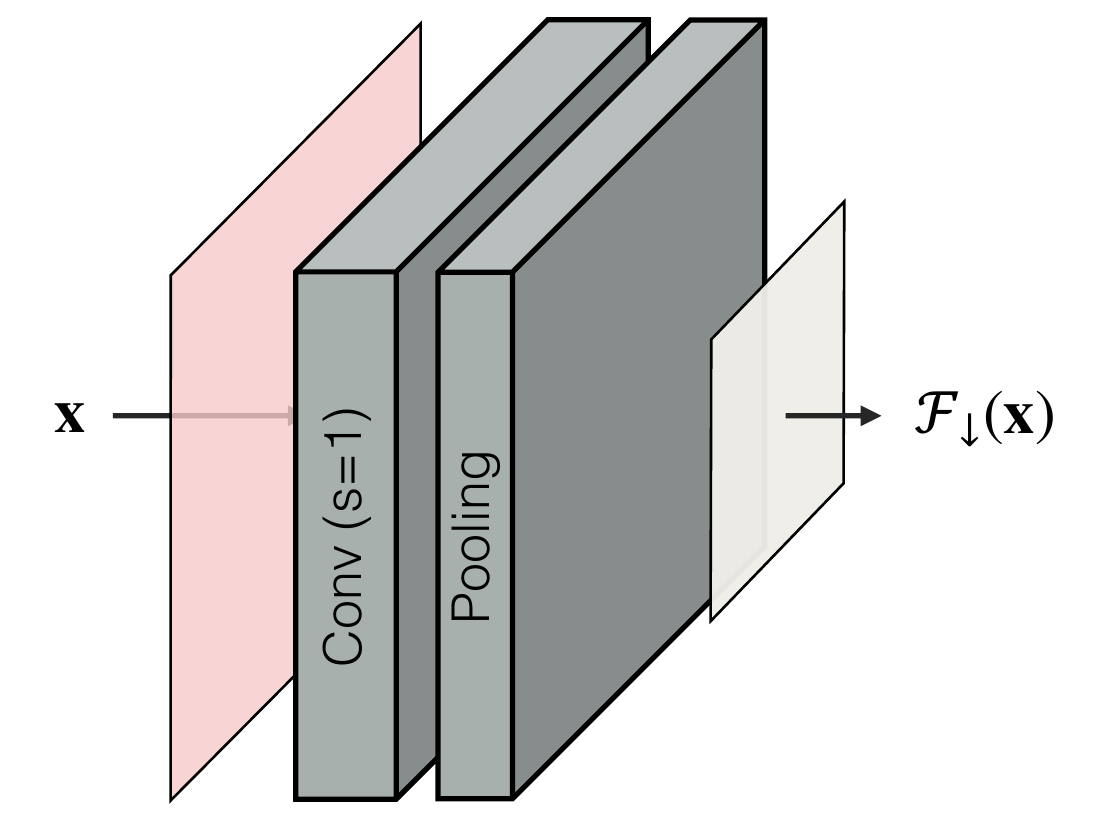}}
   \centerline{(b) CNN ($s=1$) + Pooling}\medskip
 \end{minipage}
 \hfill
 \begin{minipage}[b]{0.33\linewidth}
   \centering
   \centerline{\includegraphics[width=5.9cm]{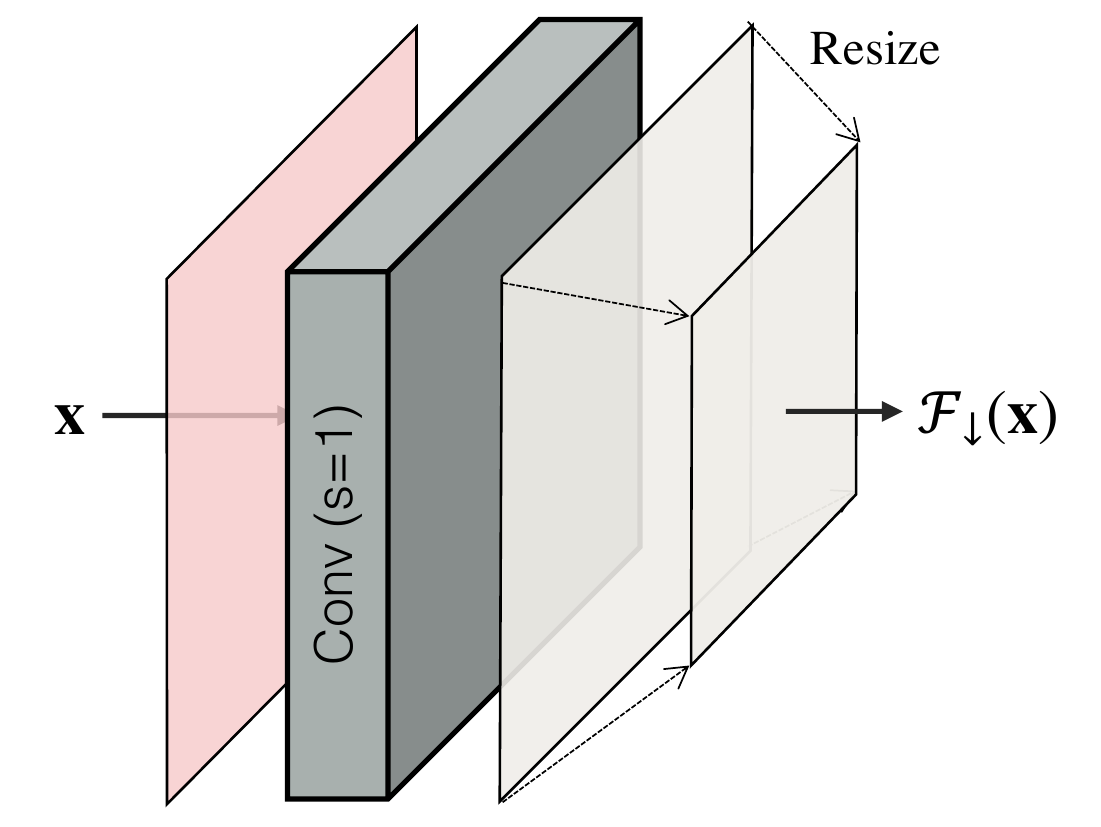}}
   \centerline{(c) CNN ($s=1$) + Resize (bilinear)}\medskip
 \end{minipage}
 \caption{Comparison of three convolutional blocks with down-sampled output. (a) A CNN block resizes by controlling the integer stride parameter $s$ (integer resizing factor). (b) A CNN block with $s=1$ resizes using an additional pooling layer (integer resizing factor) (c) Our proposed resizing module is constructed as a convolutional layer with $s=1$ followed by a resize operation, allowing for arbitrary resizing factors.}
 \label{fig_adv_example}
 \end{figure*}
Recently, deep neural networks have been applied to solve a wide diversity of video processing problems \cite{BurgerSH12, BalleLS16a, liu2019cyclicgen, SPaul2020}, including architectures designed for image resizing. Unlike legacy resizing algorithms, which rely heavily on conventional signal processing concepts, we optimize the parameters of a learned resizer in an end-to-end manner. In this way, we seek to improve adaptive streaming pipelines via CNN-based downsampling protocols (to implement the blue block in Fig. \ref{fig_abr_flow}). From a practical perspective, several design facets need consideration:
\begin{enumerate}[label= \arabic{enumi}.,ref=Step \arabic{enumi}, topsep=0pt,itemsep=-0.7ex,partopsep=1ex,parsep=1ex]
  \item Model complexity.
  \item Agnostic to video codec and upsampling algorithm.
  \item Allowing arbitrary (non-integer) scaling factors.
\end{enumerate}
Item 3. is a very important, but often neglected, feature in the design of video resizers in compression workflows. However, currently available convolutional layers can only resize their inputs by integer factors. In order to address this problem, we propose a simple alternative block, which is constructed as a convolutional layer followed by a bilinear resizer.

The outline of this paper is as follows. Section \ref{sec:related} reviews related literature, while Section \ref{sec:propose} presents details of the proposed convolutional block that allows non-integer resizing factors. Experiments and analysis are presented in Section 4, and finally, we conclude the paper in Section 5.

\section{Related Work}
\label{sec:related}
\textbf{Resizing Algorithms.} Beyond early approaches, such as bilinear, bicubic, and Lanczos, a variety of models for image and video scaling have been proposed. More recently, patch-based methods have been proposed that exploit intra \cite{Glasner2009,Freedman2011,Singh2015} or inter \cite{Freeman2002,HongChang2004,KwangInKim2010} similarities, while CNN-based models \cite{Dong2016,Wang2015,Kim2016,Tai2017,WangSR2020} have produced excellent results. However, these kinds of upscaling models are implemented on the device side, making them not inherently controllable components to the encoding side. Thus, they are of less interest in our context, since, their usability is limited in streaming workflows.

Optimized approaches to resolution reduction include better aligning local image features \cite{Kopf2013}, or preserving perceptually important details \cite{ztireli2015,Weber2016,LiuL02018}. More recently, deep learning based downsampling models have included CNN-CR \cite{LiCR2019}, which applies a 10-layer CNN to learn residuals on top of bicubic downsampled images, and a content adaptive resampler (CAR) model which estimates resampling kernels \cite{Sun2020}.

\noindent\textbf{Video Quality Assessment.} Another topic relevant to adaptive video streaming is the prediction of \textit{perceptual} quality. PSNR has been shown to be inconsistent to human perception \cite{ZhouWangLOVE2009}, especially in measuring specific distortions, such as scaling artifacts. Fortunately, many video quality models \cite{MOVIE2010,Vu2011,Pinson2014,BampisSpeed2017,tu2020bband,ztuugcvqa2020} have been proposed in recent years. In particular, SSIM \cite{WangVQA2004} and VMAF \cite{ZliVMAF16} have been very widely deployed to optimize a large fraction of compressed Internet video traffic, including downsampling conducted as part of compression.

\section{Learning a Downsampler for Adaptive Video Streaming}
\label{sec:propose}
\subsection{Proposed Convolutional Block}
We begin by comparing two commonly used convolutional blocks, shown in Figs. \ref{fig_adv_example}(a) and \ref{fig_adv_example}(b). Given an input $\mathbf{x}$ with spatial resolution $\mathrm{W}\times\mathrm{H}$, a convolutional block has trainable hyper parameters outputs $\mathcal{F}_{\downarrow\mathit{M}}(\mathbf{x})$ yielding a reduced resolution $\mathrm{W}/\mathit{M}\times\mathrm{H}/\mathit{M}$. Normally, the downsample operation by a factor $\mathit{M}$ is done by (Fig. \ref{fig_adv_example}(a))
\begin{equation}
  \mathcal{F}_{\downarrow\mathit{M}}(\mathbf{x})=\mathcal{C}_{s=\mathit{M}}(\mathbf{x}),
\end{equation}
where $\mathcal{C}_s$ denotes a convolutional operation with a \texttt{stride} parameter $s$. Alternatively, pooling layer $\mathcal{P}_{\downarrow\mathit{M}}$, typically using max pooling or average pooling, is used to reduce resolution (Fig. \ref{fig_adv_example}(b))
\begin{equation}
  \mathcal{F}_{\downarrow\mathit{M}}(\mathbf{x})=\mathcal{P}_{\downarrow\mathit{M}}\left[\mathcal{C}_{s=1}(\mathbf{x})\right].
\end{equation}

Unfortunately, the two aforementioned blocks only allow for \textit{integer} scaling factors $\mathit{M}\in\mathbb{Z}_{+}$, which may limit needed flexibility when implementing resolution changes in broader applications. To address this problem, we instead replace $\mathcal{P}$ in (2) by a differentiable resizer $\mathcal{R}$ that supports arbitrary scaling factors $\mathit{M}\in\mathbb{Q}_{+}$ (Fig. \ref{fig_adv_example}(c))
\begin{equation}
  \mathcal{F}_{\downarrow\mathit{M}}(\mathbf{x})=\mathcal{R}_{\downarrow\mathit{M}}\left[\mathcal{C}_{s=1}(\mathbf{x})\right],
\end{equation}
which we dub the \texttt{conv-resize} block. We realize the proposed building block as a convolutional layer \texttt{conv} followed by a \texttt{bilinear} resizer. Of course, the entire network can be trained end-to-end by back-propagating through the forward model, and which can be easily implemented using common libraries (e.g., TensorFlow or PyTorch). Interestingly, a similar ``\texttt{resize-conv}'' block has been used to mitigate artifacts arised from uneven overlapped responses in transposed convolution \cite{Odena16}. However, this concept has not been extended to solve the problem we discuss here.

\begin{figure}[t]
  \footnotesize
 \begin{minipage}[b]{1.0\linewidth}
   \centering
   \centerline{\includegraphics[width=8.5cm]{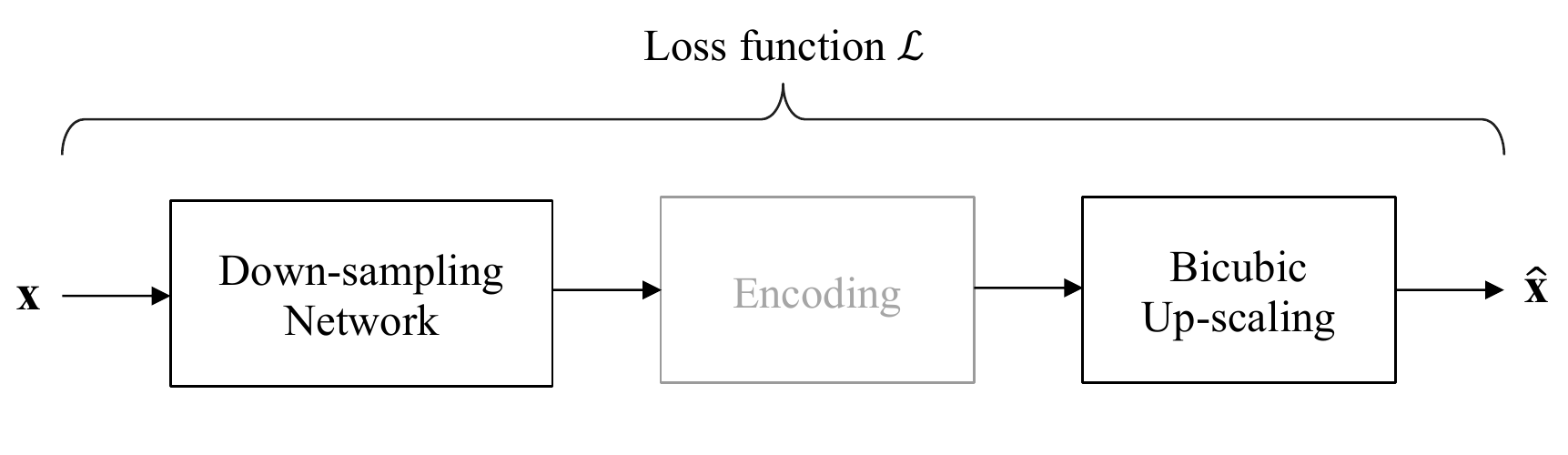}}
   \centerline{(a) Training framework. The gray block is not present in training.}\medskip
 \end{minipage}

 \begin{minipage}[b]{1.0\linewidth}
   \centering
   \centerline{\includegraphics[width=8.2cm]{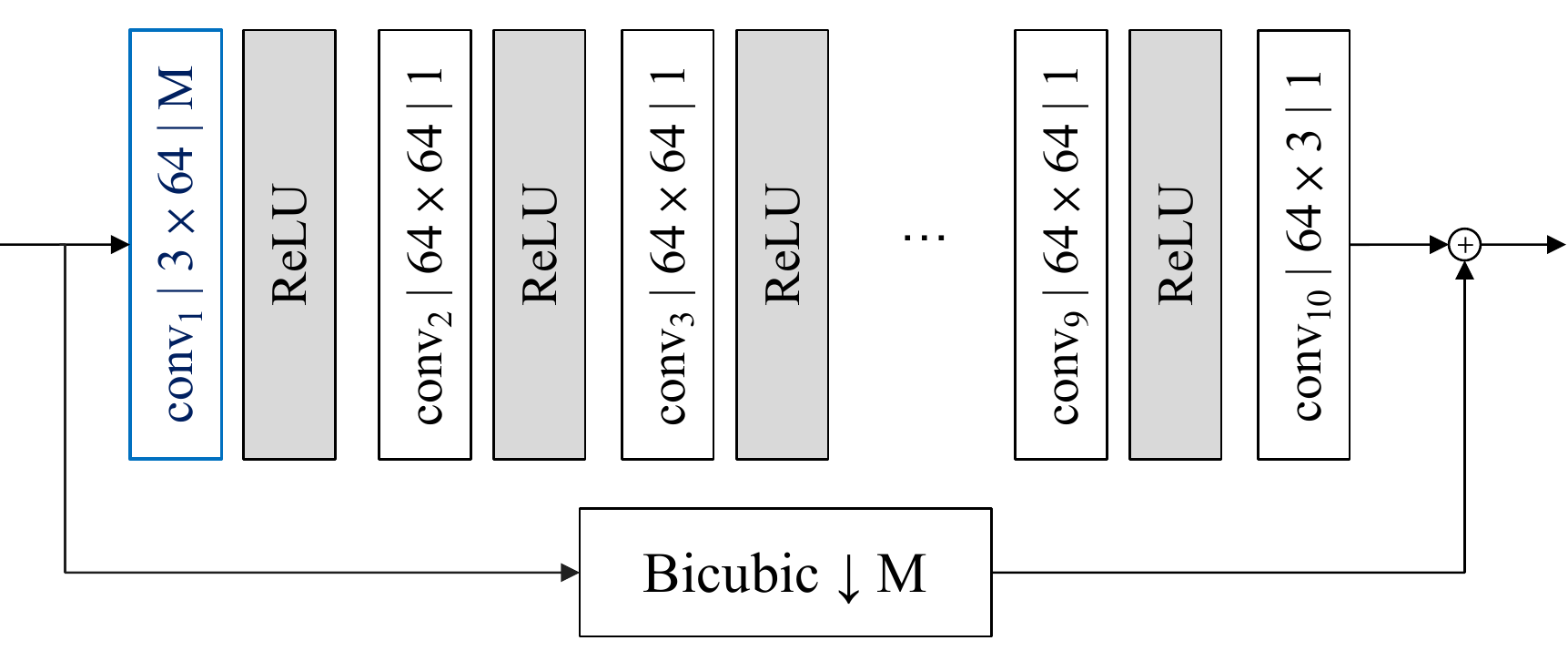}}
   \centerline{(b) Network architecture of CNN-CR \cite{LiCR2019}.}\medskip
 \end{minipage}

 \caption{Detailed (a) training framework of a downsampling network for adaptive video streaming, and (b) CNN-CR network architecture. The convolutional parameters are denoted as: input channel $\times$ output channel $\mid$ stride. Here, we replace the first convolutional layer, highlighted in blue, by a \texttt{conv-resize} layer.}
 \label{fig_train_network}
 \end{figure}

\subsection{End-to-End Training of Downsampling Network}
\noindent\textbf{Training Framework.} Aiming to end-to-end optimize the downsampling network for easy insertion into adaptive streaming scenarios, we constructed a training framework similar to the encoding pipeline in Fig. \ref{fig_abr_flow}. As shown in Fig. \ref{fig_train_network}(a), the input training data is down-scaled by a network with trainable parameters, encoded, and reconstructed to the original resolution by an upscaler. Since all conventional hybrid video encoders, such as H.264, are not \textit{differentiable} components, they are not feasible for back-propagation. Thus, we relax this problem by simply removing the encoder from the pipeline. We implemented the upscaling algorithm using bicubic interpolation as a generic and reasonably high-performance comparison. As we will show, this design achieves consistent performance gain for all the widely adopted upscaling algorithms.

Given an input source batch $\mathbf{x}$ and a reconstructed batch $\hat{\mathbf{x}}$ presented in RGB color space, the loss function is defined as the residual between $\mathbf{x}$ and $\hat{\mathbf{x}}$ of a distance function $d$:
\begin{equation}
  \mathcal{L}=d\left(\mathbf{x}, \hat{\mathbf{x}}\right)=\norm{\mathbf{x}-\hat{\mathbf{x}}}_2^2.
\end{equation}
Here, we use the mean squared error loss ($d(x)=\norm{x}_2^2$), which maximizes the PSNR of the reconstructed images. 

\noindent\textbf{Network Architecture.} We use the previously mentioned CNN-CR model as the backbone of the downsampling network. As shown in Fig. \ref{fig_train_network}(b), the 3-channel RGB signal is fed into the network. The architecture of CNN-CR consists of $10$ stages of convolutional layers. The sizes of the convolution kernels are fixed at $3\times 3$ in all the layers, while the number of filters is $64$ for all of the first $9$ stages. Except for the $\mathrm{conv_{10}}$ layer, all of the convolutional layers are activated by a ReLU nonlinearity. Finally, a 3-channel output is produced having reduced size, yielding a residual added to the bicubic down-sampled input image. The parameterization of each layer is detailed in the figure. It should be note that the subsampling process happens in the first convolution layer, by the stride $s=\mathit{M}$. To remove the limitation of an integer $\mathit{M}$, the first layer $\mathrm{conv_1}$ is replaced by our proposed \texttt{conv-resize} block.

\section{EXPERIMENTS}
\label{sec:experiments}

\subsection{Implementation Details and Experimental Setup}
We used the TensorFlow framework (version 1.15) to implement the proposed non-integer deep downsampling method. The Adam solver \cite{kingma:adam} was used to optimize the networks, with parameters $(\beta_1, \beta_2)=(0.9,0.999)$ and a batch size of $16$. The networks were trained on 500K iterations of back-propagation, with a learning rate that was fixed at $1e-4$. The training images were randomly cropped to $M\lfloor\frac{256}{M}\rfloor\times M\lfloor\frac{256}{M}\rfloor$, which is divisible by the scaling factor $\mathit{M}$. 

We used DIV2K \cite{Agustsson2017}, an image dataset consisting of $1000$ very high quality pictures,\footnote[1]{https://data.vision.ee.ethz.ch/cvl/DIV2K/} as training data. This dataset was designed for studying image super-resolution problems. All of the images in it have 2K pixels along either the vertical or horizontal axis. To evaluate our method under the adaptive video streaming scenario, we utilized $20$ test video contents of 1080p resolution and YUV420 format obtained from Xiph Video Test Media,\footnote[2]{https://media.xiph.org/video/derf/} We also used $25$ 1080p video sources collected from the Netflix library, yielding more diverse and realistic contents. To integrate the network into a video encoding pipeline, the video formats input to the network were in RGB888 format. Following resolution reduction, the videos were converted back to their original format (YUV420) prior to encoding.

\begin{table*}[!t]
  \small
  \renewcommand{\arraystretch}{1.2}
  \centering
  \renewcommand{\tabcolsep}{4.5pt} 
  \begin{tabular}{ l || ccc || ccc || ccc || ccc }
  \hline
  Upsampler     && \multicolumn{4}{c}{bilinear$\uparrow$} &  
                && \multicolumn{4}{c}{bicubic$\uparrow$} &   \\
                \cline{1-1} \cline{2-4} \cline{5-7} \cline{8-10} \cline{11-13}
  Downsampler    & \multicolumn{3}{c||}{CNN-CR$\downarrow$}  
                 & \multicolumn{3}{c||}{Proposed$\downarrow$}
                 & \multicolumn{3}{c||}{CNN-CR$\downarrow$}  
                 & \multicolumn{3}{c}{Proposed$\downarrow$} \\
                 \cline{1-1} \cline{2-4} \cline{5-7} \cline{8-10} \cline{11-13}
  BD-rate metric &~PSNR~ & SSIM~ & VMAF~ 
                 &~PSNR~ & SSIM~ & VMAF~ 
                 &~PSNR~ & SSIM~ & VMAF~
                 &~PSNR~ & SSIM~ & VMAF~ \\
  \hline
  $\mathit{M}=1.5$      & ---        & ---        & --- 
                        & \phZ-4.06  & \phZ-2.47  & \phZ-1.20 
                        & ---        & ---        & --- 
                        & \phZ-2.22  & \phZ-1.19  & \phZ-0.77 \\
  $\mathit{M}=2$        & \phZ-4.84  & \phZ-4.90  & \phZ-2.45
                        & \phZ-4.72  & \phZ-4.74  & \phZ-2.28
                        & \phZ-2.68  & \phZ-3.18  & \phZ-2.16  
                        & \phZ-2.64  & \phZ-3.11  & \phZ-2.11 \\
  $\mathit{M}=2.5$      & ---        & ---        & --- 
                        & \phZ-4.59  & \phZ-6.54  & \phZ-6.88 
                        & ---        & ---        & --- 
                        & \phZ-2.56  & \phZ-4.53  & \phZ-5.34 \\
  $\mathit{M}=3$        & \phZ-4.27  & \phZ-8.00  & -10.51 
                        & \phZ-4.51  & \phZ-8.17  & -11.26
                        & \phZ-2.86  & \phZ-6.33  & \phZ-8.78 
                        & \phZ-2.77  & \phZ-6.55  & \phZ-9.25  \\
  $\mathit{M}=4$        & \phZ-4.06  & -10.36     & -14.70 
                        & \phZ-5.48  & -11.88     & -15.97 
                        & \phZ-1.31  & \phZ-7.42  & -12.99 
                        & \phZ-3.11  & \phZ-9.14  & -15.12 \\
  $\mathit{M}=5$        & \phZ-2.01  & \phZ-9.72  & -22.37 
                        & \phZ-3.28  & -11.37     & -24.17 
                        & \phZ+0.52  & \phZ-6.62  & -19.24
                        & \phZ+0.16  &\phZ -7.74  & -20.20\\

  \hline
  \end{tabular}
  \caption{Comparison of the performances of a conventional convolutional layer and the proposed \texttt{conv-resize} layer deployed in CNN-CR: average change of BD-rate expressed as percentage. The baseline of comparison is the Lanczos downsampling algorithm with the same encoding recipe. Smaller or negative values indicate better coding efficiency. A ``---'' in a cell indicates the result is not applicable to the model.}
  \label{tab:comparison}
\end{table*}

\begin{table}[!t]
  \small
  \renewcommand{\arraystretch}{1.2}
  \centering
  \renewcommand{\tabcolsep}{3.4pt} 
  \begin{tabular}{ l | c | c | c | c | c | c }
  \hline
  Scaling factor $\mathit{M}$ & $1.5$ & $2$ & $2.5$ & $3$ & $4$ & $5$\\
  \hline
  \texttt{conv-resize} & -2.22 & -2.64 & -2.56 & -2.77 & -3.11 & +0.16\\
  \texttt{resize-conv} & -0.98 & -1.46 & +6.73 & +17.27 & +18.88 & +31.51 \\
  \hline
  \end{tabular}
  \caption{Comparison of using \texttt{conv-resize} and \texttt{resize-conv} as the first layer of CNN-CR$\downarrow$ with bicubic$\uparrow$. Each cell presents the average PSNR BD-rate expressed as percentage.}
  \label{tab:study_order}
  \end{table}

\subsection{Quantitative Comparison}
We measured the objective coding efficiency of each downsampling model within the same video encoding pipeline using the Bj\o{}ntegaard-Delta bitrate (BD-rate) \cite{BDRate01}, which quantifies average differences in bitrate at the same distortion level relative to another reference encode. To calculate BD-rate, we encoded the down-sampled videos by x264 at $15$ different QPs, ranging from $17$ to $46$. Then, the encoded videos were up-scaled back to their original resolutions. The performances of all of the downsamplers were compared to the same baseline - the Lanczos downsampling algorithm implemented in ffmpeg. A negative number of BD-rate means the bitrate was reduced as compared with the baseline. Lastly, the distortion levels that were used for BD-rate calculation were quantified using PSNR, SSIM, and VMAF.

The performance results are shown in Table~\ref{tab:comparison}, with respect to different objective video quality models. We report the BD-rate changes obtained relative to the baseline (Lanczos downsampling under the same conditions), averaged over all the videos in the test set. We comprehensively evaluated the proposed downsampling networks for various scaling factors that are commonly used in practice, using two different upsampling algorithms. From the results in the table, we can draw a number of conclusions. First, these results show that learned downsampling models are able to further optimize existing video encoding pipelines. Indeed, significant BD-rate reductions were obtained in many cases. The proposed \texttt{conv-resize} block performed closely to the conventional block at integer scaling factors, and yielded reasonable BD-rates when $\mathit{M}$ was not an integer. It may also be observed that, when using bicubic upsampling when $\mathit{M}=5$, the PSNR BD-rate was slightly worse than the baseline scenario. It is possible that the models were trained without considering encoding effects, such as distortions and rate consumption, hence resulting in a suboptimal result. However, the subtle suboptimality in PSNR BD-rate is negligible, since PSNR is not the ultimate optimization target.

Despite training with a fixed bicubic upsampler, the models were still able to generalize well to bilinear upsampling. In fact, the performance results reveal similar trends on the two upsampling algorithms. However, more significant BD-rate improvements were obtained on bilinear upsampling, which usually results in worse quality, leaving greater room for improvement of objective video quality. We have also observed very similar results using other upscaling algorithms, including Lanczos. Another interesting observation that can be made is that the CNN-based models delivered coding gains with respect to all of the BD-rate measurements.

\subsection{Which Goes First? Convolution or Resizing?}
By combining the downsampling process with a resizer, non-integer scaling is preserved if the order of \texttt{conv} and \texttt{resize} is reversed. To study this alternate model, we compared two blocks using both orderings in Table \ref{tab:study_order}. It may be observed that, on all the scaling factors, the proposed block \texttt{conv-resize} delivered significantly better results than \texttt{resize-conv} on the task of video downsampling. This is not surprising, since the interpolation video signals were decimated, with loss of information, at the input of the network. Placing a learnable convolutional layer beforehand allows the retention of information, yielding better performance. This is quite analogous to the classic signal processing rate change structure of an anti-aliasing filter placed prior to a decimator. We also observed that the performance of \texttt{resize-conv} dropped sharply with increases of $\mathit{M}$, as a consequence of more severe information loss.


\section{Conclusion}
\label{sec:conclusion}
We designed a convolutional block that allows fractional resizing in downsampling networks. The proposed block is simple and can be effectively implemented within diverse deep learning platforms. We believe that the new block can also be applied in different types of downsampling architectures. Looking further ahead, we plan to extend the idea to other important problems, such as the quality assessment of pictures of different scales, or viewed at different distances.

\pagebreak
\section{References}
\bibliographystyle{IEEEtran}
{\footnotesize\bibliography{strings}}

\end{document}